\documentclass[11pt,A4]{revtex4}
\usepackage{amsmath}
\usepackage{graphicx}
\usepackage{subfigure}
\usepackage{color,soul}

\begin{document}

\title{An integral fluctuation theorem for systems with unidirectional
transitions}

\author{Saar Rahav$\;^1$}
\author{Upendra Harbola$\;^2$}
\affiliation{$\;^1$Schulich Faculty of Chemistry, Israel Institute of
Technology, Haifa 32000, Israel.}
\affiliation{$\;^2$Department of inorganic and physical chemistry, Indian
Institute of Science, Bangalore, 560012, India.}

\begin{abstract}
The fluctuations of a Markovian jump process with one or more unidirectional transitions,
where $R_{ij} >0$ but $R_{ji} =0$, are studied. We find that such systems
satisfy an integral fluctuation theorem. The fluctuating quantity satisfying the theorem
is a sum of the entropy produced in the bidirectional transitions and a dynamical 
contribution which depends on the residence times in the states connected by the unidirectional 
transitions. The convergence of the integral fluctuation theorem is studied numerically, and found to
show the same qualitative features as in systems exhibiting microreversibility.
\end{abstract}

\maketitle
\section{Introduction}

The last two decades have seen substantial advancement in our understanding
of the thermodynamics
of small out-of-equilibrium systems. Much of the progress was related to the
study of fluctuations in such systems.
In particular, it was found that many out-of-equilibrium processes satisfy
fluctuation
theorems~\cite{Evans1993,Evans1994,Gallavotti1995,Kurchan1998,Lebowitz1999,
Seifert2005, Rahav2013,Harbola2014}. These celebrated relations compare
the probabilities to observe a realization of a process and its
time-reversed symmetry related counterpart. 
The ratio of these probabilities is expressed in terms of
thermodynamic quantities such as entropy production
or heat.
A closely related set of results is termed work
relations~\cite{Jarzynski1997,Crooks1998}.
The latter focus on the fluctuations in the work done on the system when it is
driven away
from equilibrium.

Fluctuation theorems are valid for systems which are driven arbitrarily far from
their thermal equilibrium. 
The discovery of fluctuation theorem has opened up new research directions and enhanced our
qualitative and quantitative
understanding of small systems in contact with thermal environments. Some of
the progress made
is summarized in several review
articles~\cite{Evans2002,Maes2003,Bustamante2005,Kurchan2007,Harris2007,
Imparato2007,Ritort2008,Esposito2009,Broeck2010,Ciliberto2010,Jarzynski2011,
Campisi2011,Seifert2012}.

One of the important concepts underlying fluctuation theorems is the ability to
meaningfully and consistently
assign thermodynamic interpretation to
a single realization of a stochastic out-of-equilibrium process. This approach
is sometimes referred to as stochastic thermodynamics. Sekimoto has
demonstrated that 
heat and work can be defined for a single realization of a process so that the first law is
satisfied~\cite{Sekimoto1997}.
Seifert has introduced the concept of a fluctuating system entropy and demonstrated that it allows to obtain an 
exact fluctuation theorem~\cite{Seifert2005}. Fluctuation theorems can be viewed as
replacing the inequality of the second law of thermodynamics by
an equality for the exponential average of a realization dependent fluctuating
quantity~\cite{Jarzynski2011}. 
The inequality is restored for the ensemble average of this quantity with the
help of the Jensen inequality.
More information about stochastic thermodynamics can be found in
Seifert's comprehensive review~\cite{Seifert2012}.

Derivations of fluctuation theorems commonly use a direct comparison of the
probabilities of a realization and \textcolor{red}{that} of its time reversed counterpart. 
For systems driven by a given force protocol,
one similarly compares probabilities of realization of 
a forward process to that of
its time reversed (backward) process driven by a time reversed force protocol.
The same considerations can be applied for systems in absence of time reversal
symmetry, for instance 
due to the presence of a magnetic field, see Ref. \cite{Wang2014} for a recent
review.
All of those approaches employ microreversibility, namely the fact that
time inversion (combined with an inversion of momenta, driving protocol, and
possibly magnetic field)
maps between allowed realizations of a forward and a backward process. In jump
processes microreversibility means that if the transition from
state $i$ to state $j$ has a finite rate, $R_{ji} >0$, then so does the reversed
transition, $R_{ij} >0$, where we assumed absence of magnetic fields.

There are instances however where the simplest description of a natural
process is one where microreversibility is violated. Consider an atom
in an excited state which decays via spontaneous emission of a photon which escapes from the system.
In many situations it is useful to derive a reduced description
for the atom in which the field serves as an external reservoir.
For spontaneous emission the empty field modes can be interpreted as a zero temperature 
reservoir. The reversed process, namely an excitation of the atom, requires 
presence of photons, but when there are no such photons, this reverse process will not occur. 
As a result the reduced description has
 $R_{\text{emission}} >0$ while $R_{\text{absorption}} =0$. When the field modes are in a thermal
state with finite temperature the presence of stimulated processes restore microreversiblity.
Unidirectional transitions can be incorporated into models
of heat engines and machines
which also include reversible transitions, such as the model of a photosynthetic reaction center
studied by Dorfman et. al. \cite{Dorfman2012} (See Fig. 2 there). Irreversible jump processes are also used to model 
biological enzymes which break down the substrate they move on, such as cellulase, see e.g. \cite{Ting2009}.
In the following we use this as motivation to study jump processes 
which violate microreversibility. We focus on the fluctuations in such systems
and ask whether they satisfy a fluctuation theorem.

 Fluctuations of systems with unidirectional transitions have generated limited interest so
far.
Ohkubo derived a fluctuation theorem which holds also for irreversible systems \cite{Ohkubo2009}. It
is based on a posterior transition rates obtained with the help of 
Bayes' theorem.
A fluctuation theorem for a system of soft spheres with dissipative 
collisions was derived by Chong et. al. \cite{Chong2010}. It applies to systems
with continuous dynamics and moreover requires thermal initial conditions.
Other approaches used to investigate fluctuation theorems in
systems with an irreversible transition replace the vanishing rate by an effective finite rate using some coarse-graining of
the dynamics.
 Ben-Avraham, {\em et. al.} 
have suggested to measure the state of the system in fixed time intervals
\cite{Avraham2011}. This coarse graining in time allows for an effective backward rate which is actually obtained
from the combined contributions
 of allowed transitions which take the system to the other side of the irreversible transition.
Zeraati {\em et. al.} chose to view the vanishing transition rate
as being the limit of a very small but finite rate, which is small enough to be 
unlikely to be observed in a finite time experiment~\cite{Zeraati2012}. This effective rate was then
estimated using Bayes theorem which in turn is then used to obtain a lower bound 
for the entropy production that depends on the observation time. 
Both of those approaches exhibit logarithmically diverging quantities which are ill defined in the limit where 
the coarse-graining is removed, namely for vanishing time intervals between measurements or infinite observation time.

Here we employ an approach which does not suffer from such difficulties, and show that an integral fluctuation
theorem
holds for systems with unidirectional jumps. This fluctuation theorem is based on a different
treatment of reversible and irreversible transitions. It holds for a fluctuating quantity
which is a sum of two contributions. The first is the usual 
fluctuating entropy production due to reversible transitions.
The second is an unusual dynamical term which depends on
the fluctuating residence times in the states connected by the irreversible
(unidirectional) transitions. This prescription avoids the difficulties in defining an entropy production
for the irreversible terms that led to diverging expressions in the coarse-grained approaches.

The structure of the paper is as follows. In Sec. \ref{sec:4states}
we consider a simple example of a jump process with a single irreversible transition
and
derive the integral fluctuation theorem. The derivation is simple and can be
easily applied to
systems with more states, irreversible transitions, or time dependent transition
rates. Such generalizations are straightforward, and
are stated without detailed proof in Sec. \ref{sec:gen}. In Sec. \ref{sec:conv}
we discuss the number of realizations needed
for convergence of the exponential average appearing in the integral fluctuation
theorem. We point out that estimates based on the
identification of typical and dominant realizations which were developed for
systems with reversible rates
are also applicable here. We
summarize our results in Sec. \ref{sec:disc}.

\section{A simple model}
\label{sec:4states}

We introduce the integral fluctuation theorem with the help
of a simple example of a Markovian jump process. The use of an example allows
us to present the derivation without using unnecessarily complicated notations.
The choice of the example is based on two requirements. We want the system to 
include one irreversible transition. In addition, we want at least one closed cycle
of reversible transitions, to allow for a steady state flux even in the absence of the irreversible transition.
These considerations lead us to study a system with four states and one
irreversible transition, which is the minimal model that has no more that one transition between states and satisfies the requirements. 
Generalizations to more general jump processes are possible and will be described In Sec. \ref{sec:gen}.
 The possible transitions between the four states are
characterized 
by the transition rates, with $R_{ij} \ge 0 $ corresponding to the transition $j \rightarrow i $ (for $i \ne j$).
When the transition between $i$ and $j$ is reversible the combination
$\ln \frac{R_{ij}}{R_{ji}}$ is interpreted as the entropy change in the
reservoir
during the transition \cite{Lebowitz1999}. This identification is motivated by
the fact that for thermally activated rates this term 
commonly has the form $\frac{E_{i}-E_{j}}{T}$, with $E_{i}$ the energy of state
$i$ and 
$T$ the temperature of the reservoir.
This is an ill-defined quantity for unidirectional rates, where $R_{ij} > 0$
and $R_{ji} = 0$.

\begin{figure}
\begin{center}
\includegraphics[scale=0.4]{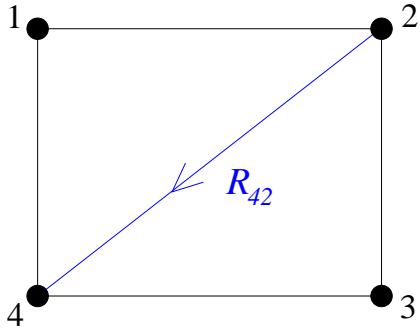}
\caption{A graph representation of the jump process with one unidirectional
transition studied here. \label{fig:system}}
\end{center}
\end{figure}

The simple model investigated in this section can be conveniently represented
using a graph, which is depicted in Fig. \ref{fig:system}.
The black solid lines represent reversible (bidirectional) transitions. In contrast,
the transition from state $2$ to state $4$ is irreversible, with rates $R_{42} >0$ and $R_{24}=0$.

The system's probability distribution evolves according to a master equation,
$\frac{d \mathbf{p}}{dt}= \mathbf{R} \mathbf{p}$,
with $R_{ii} \equiv - \sum_{j \ne i} R_{ji} = - r_{i}$. 
Here $\mathbf{P}$ is the vector containing population of the four states and 
$\mathbf{R}$ is the transition rate matrix.
If left alone the system
relaxes to a steady state, $\mathbf{p}^{ss}$.
A history, or equivalently a realization, of the system is a list detailing the
state of the system at any given time including also the specific transitions it
made during the entire realization. We denote a given realization by $\gamma$. For example
\begin{equation}
\label{eq:frealization}
\gamma =\left\{ 2 \xrightarrow{t_{1}} 3 \xrightarrow{t_{2}} \cdots 4
\xrightarrow{t_{n}} 1 \right\}
\end{equation}
corresponds to a realization where the system was initially at state $2$,
and stayed there until time $t_{1}$. (This can be denoted equivalently by
$\gamma (t) =2$ for $0 \le t < t_{1}$.)
At time $t_{1}$ the system makes a transition to state $3$, etc. Eventually the
system makes a transition from state $4$ to state $1$ at time
$t_{n}$. The system then stays in that state until the end of the observation at
time $t_{f}$.
Since this is a Markovian jump
process the probability density of such a history is
\begin{equation}
P(\gamma)= p_{i}(2) e^{-r_{2}t_{1} } R_{32} e^{-r_{3}(t_{2}- t_{1}) } \cdots
R_{14}  e^{-r_{1}(t_{f} - t_{n}) },
\end{equation}
where $\mathbf{p}_{i}$ denotes the initial probability distribution. We will
denote by $\mathbf{p}_{f}$
the final probability distribution, namely the solution of the master equation
at time $t_{f}$, given the initial 
condition $\mathbf{p}_{i}$.

Let us denote by $\overline{\gamma}$ the time reversed realization of $\gamma$.
For $\gamma$ of Eq. (\ref{eq:frealization}) this
is clearly
\begin{equation}
\label{eq:brealization}
\overline{\gamma} = \left\{ 1 \xrightarrow{t_{f}- t_{n}} 4 \cdots
\xrightarrow{t_{f} - t_{2}}  3 \xrightarrow{ t_{f} - t_{1}} 2  \right\}.
\end{equation}
We note that time reversal is a one-to-one mapping between realizations.
However, many of the time reversed realizations are not allowed 
under the dynamics of the irreversible jump process depicted in
Fig.~\ref{fig:system} because they would
make the forbidden $4 \rightarrow 2$ transition. We note that there are
approaches that derive integral fluctuation theorems by separating realizations
into
groups of regular and irregular realizations and calculating the weight of the
latter~\cite{Murashita2014}, but a simpler approach is possible here. We view
the reversed realizations
$\overline{\gamma}$ as obtained from an auxiliary dynamics in which one reverses
the direction of the irreversible transition. For the simple 
example considered in this section this auxiliary dynamics has
$\overline{R}_{24}={R}_{42}$ and $\overline{R}_{42}=0$ and otherwise
$\overline{R}_{ij}={R}_{ij}$ (for $i\ne j$). We intentionally avoid the more
common terminology of "forward" and "backward"
processes, which are best left for cases in which the backward process has a meaningful 
physical interpretation.

Importantly, the time reversal mapping between $\gamma$ in the physical dynamics
and $\overline{\gamma}$ in the auxiliary dynamics is one-to-one. One can assign
a probability density $\overline{P} (\overline{\gamma})$
for realizations of the auxiliary dynamics. For the realization in  Eq.
(\ref{eq:brealization}) one finds
\begin{equation}
\overline{P}(\overline{\gamma})= \overline{p}_{i} (1) e^{-\overline{
r}_{1}(t_{f} - t_{n} )} \overline{R}_{41}  \cdots e^{-\overline{ r}_{3}(t_{2}-
t_{1}) } \overline{R}_{23}  e^{-\overline{ r}_{2} t_{1} }.
\end{equation}
We note that $\overline{r}_{i} \ne r_{i}$ for the states linked by the
unidirectional transition. It is crucial to point out that with
the interpretation of irreversibility as coming from spontaneous emission the
auxiliary dynamics is not physical. It involves transitions
taking the system from a low energy state to an excited state without energy
input from the environment.

The auxiliary dynamics allows us
to obtain a formal integral fluctuation theorem by defining
\begin{equation}
\label{eq:defsigma}
\Sigma (\gamma) \equiv \ln \frac{P(\gamma)}{ \overline{P} (\overline{\gamma}) },
\end{equation}
and noting that
\begin{equation}
\label{eq:intft}
\left< e^{-\Sigma} \right>=\sum_{\gamma} e^{-\Sigma (\gamma)} P(\gamma) =
\sum_{\overline{\gamma}} \overline{P} (\overline{\gamma})=1.
\end{equation}
Equation (\ref{eq:intft}) is a mathematical identity. Fluctuation theorems of
this type are of interest when they can be given 
an appealing physical interpretation. We show that it is possible to express
$\Sigma (\gamma)$ using only properties
of the physical dynamics, suppressing  any explicit dependence on the auxiliary
dynamics.

To do so we note that Eq. (\ref{eq:intft}) is valid for {\em any} choice of the
initial condition for the auxiliary dynamics. We therefore choose
$\overline{\mathbf{p}}_{i} = \mathbf{p}_{f}$,
namely the initial distribution of the auxiliary dynamics is the same
as the final distribution of the physical dynamics. A short calculation then finds
\begin{equation}
\label{eq:sigma}
\Sigma (\gamma)=\Delta S_{rev} (\gamma) + R_{42} 
(\tau_{4}(\gamma)-\tau_{2}(\gamma)).
\end{equation}
Here
\begin{equation}
\label{eq:DeltaS}
\Delta S_{rev} (\gamma) = \sum_{i,\textrm{rev}} \ln \frac{R_{\gamma_{i+1}
\gamma_{i} } }{R_{\gamma_{i} \gamma_{i+1} }} + \ln \frac{p_{i} (\gamma
(0))}{p_{f} (\gamma (t_f))},
\end{equation}
where the first term on the right hand side is the sum of contributions to the
medium entropy production from all the reversible transitions during the realization
$\gamma (t)$,
while the second is the change of the fluctuating system entropy
\cite{Seifert2005}. Note that the quantity $\Sigma (\gamma)$ fluctuates
from one realization to another. 
$\tau_{i} (\gamma)$ in Eq. (\ref{eq:sigma}) denotes the (fluctuating) time that
the system spends in state $i$ during the realization.
It can be written as $\tau_{i} (\gamma) =\int dt \chi_{i} (\gamma(t))$ where
$\chi_{i}=1$ for $\gamma(t)=i$ and $0$ otherwise.

The jump process depicted in Fig.~\ref{fig:system} therefore satisfies the
integral fluctuation theorem
\begin{equation}
\label{eq:finalIFT}
\left< e^{-\Delta S_{rev} +  R_{42}(\tau_{2}-\tau_{4})} \right>=1,
\end{equation}
which is accompanied by a second law like inequality
\begin{equation}
\label{eq:ineq1}
\left< \Delta S_{rev} + R_{42} (\tau_{4}-\tau_{2}) \right> \ge 0.
\end{equation}
The quantity $\Sigma$ satisfying the
integral fluctuation theorem (\ref{eq:finalIFT}) has a simple physical interpretation which is
an interesting mixture of dynamical and thermodynamic quantities. 
The thermodynamic part, $\Delta S_{rev}$, includes the change in medium entropy in the
finite temperature reservoirs. The thermodynamic interpretation of this entropy production as resulting
from e.g. energy exchanged with finite temperature reservoirs is well understood, in contrast to the
 absence of a similar interpretation for irreversible transitions.
 The contribution of the irreversible transitions is dynamical and
depends on residence times, namely
the time that the system spends in the states connected by the transition. The
structure of this dynamical term, a product of the transition rate times
residence times, is similar to the so called traffic which contributes to the
linear response of jump processes \cite{Maes2008,Baiesi2009}. 
Both the dynamical term in Eq. (\ref{eq:finalIFT}) and the traffic
as considered in Refs. \cite{Maes2008,Baiesi2009} are given by time integrals of escape rates, but
only the irreversible transition contributes to the dynamical part of $\Sigma$
 and furthermore the sign of the contribution depends on whether the irreversible transition points into, or out of, the state.

It is worthwhile to examine the inequality (\ref{eq:ineq1}) more closely at
steady state, where all the terms appearing in the inequality are 
linearly proportional to time, $\left<\Delta S_{rev} \right>
= q_{rev} t_{f}$ and $\left< \tau_{i} \right> = p^{ss} (i)  t_{f}$. The
inequality (\ref{eq:ineq1})
can be rewritten as
\begin{equation}
\label{eq:AvS}
q_{rev} + R_{42} (p^{ss} (4) -p^{ss} (2)) \ge 0.
\end{equation}
Here $q_{rev}$ is the reversible entropy production rate at steady state.
 Each reversible transition $i \rightarrow j$ contributes
 $\mbox{ln}(R_{ji}/R_{ij})$ to the entropy production. At steady state the mean 
 rate of the $i \rightarrow j$ transition is $R_{ji} p^{ss} (i)$. Therefore
 \begin{eqnarray}
 \label{eq:EntropyRate}
 q_{rev} = \sum_{(i,j),\mathrm{rev}} \left[R_{ji} p^{ss} (i)- R_{ij} p^{ss} (j) \right] \mbox{ln}(R_{ji}/R_{ij})
 \end{eqnarray}
where the sum is over all unordered pairs of states which are connected by a reversible transition. 
(For simplicity we assume here that at most one transition connects any give pair of states.)
Finally, with the help of the conservation laws for steady state fluxes this inequality
can be recast as
\begin{multline}
\label{eq:exampleineq}
J^{ss}_{14} \ln \frac{R_{14} p^{ss} (4)}{ R_{41} p^{ss} (1)}+J^{ss}_{21} \ln
\frac{R_{21} p^{ss} (1)}{ R_{12} p^{ss} (2)}+J^{ss}_{34} \ln \frac{R_{34} p^{ss}
(4)}{ R_{43} p^{ss} (3)}
+J^{ss}_{23} \ln \frac{R_{23} p^{ss} (3)}{ R_{32} p^{ss} (2)} \\ + R_{42} p^{ss}
(2) \left[\ln \frac{p^{ss} (2) }{p^{ss} (4) } + \frac{p^{ss} (4) }{p^{ss} (2) }
- 1 \right] \ge 0.
\end{multline}
The terms of the form $J^{ss}_{ij} \ln \frac{R_{ij} p^{ss} (j)}{ R_{ji} p^{ss}
(i)} = \left(R_{ij} p^{ss} (j)- R_{ji} p^{ss} (i) \right)  \ln \frac{R_{ij}
p^{ss} (j)}{ R_{ji} p^{ss} (i)} \ge 0$ commonly appear in systems with
reversible transitions and are interpreted as a product of flux and
thermodynamic affinity. Interestingly, the presence of a unidirectional transition leads to the
appearance of a different type of term in Eq. (\ref{eq:exampleineq}), namely the last
term on the left hand side. This term is also non negative since $\ln x +
\frac{1}{x} - 1 \ge 0$ for positive $x$. $R_{42} p^{ss}(2)$ is clearly the steady state 
flux of irreversible transitions. However, it is not clear whether $\ln \frac{p^{ss} (2) }{p^{ss} (4) } + \frac{p^{ss} (4) }{p^{ss} (2) }
- 1  = \ln \frac{ \left< \tau_{2} \right>}{\left< \tau_{4} \right> } + \frac{\left< \tau_{4} \right> }{\left< \tau_{2} \right> }- 1, $
which depends on the ratio of likelihood to find the system at both sides of the unidirectional transition,
 can be
meaningfully interpreted as some generalized affinity. 

\section{Possible generalizations}
\label{sec:gen}

In this section we briefly describe various generalizations of the integral 
fluctuation theorem, Eq. (\ref{eq:finalIFT}).
The derivations are straightforward and most of the details are omitted.

We first note that there is some mathematical freedom in the choice of possible auxiliary
dynamics. 
One can modify the magnitude of various transition rates in
the auxiliary dynamics, and Eq. (\ref{eq:intft}) will still hold, as long
as the correct transitions are prohibited.
However, this freedom to play with the magnitude of rates results in a $\Sigma$
whose "entropic" and "dynamical" parts have dubious physical interpretation.
For instance, using an auxiliary dynamics with a modified value of
$\overline{R}_{24}$ would result in contributions of $\ln \frac{{R}_{42}}{\overline{R}_{24}}$
to the "entropy production". But such a physical interpretation is unjustified since the rate $\overline{R}_{24}$
has nothing to do with the dynamics of the physical system. 
Moreover, with this choice of auxiliary dynamics $\Sigma$ depends on
the rate $\overline{R}_{24}$, and the resulting integral fluctuation theorem no longer depends only
on properties of the physical dynamics.
The auxiliary dynamics used in 
Sec. \ref{sec:4states} was chosen to prevent the appearance
of such difficulties, and keep the physical interpretation of $\Sigma$ transparent. 
In that sense the demand for consistent physical interpretation suggests that
the auxiliary dynamics should be the one that was used in Sec. \ref{sec:4states}.

The derivation of Eq. (\ref{eq:DeltaS}) presented in Sec. \ref{sec:4states} never made use of the fact
that there are only four states 
in the system. It applies to a jump process with any finite number of states as long as the initial
and final probability distributions have finite values for 
all states. Even when some of the probabilities $p_{f}, p_{i}$ vanish,
it is possible that the approach developed
by Murashita {\em et. al.} \cite{Murashita2014} may be of use, but this is
beyond the scope of the current paper.

Another possible generalization is to a system with several unidirectional
transitions. In this case
the auxiliary dynamics is one where all the irreversible rates have been
reversed. Their contribution
to the fluctuation theorem enters through the escape rates
$r_{i}$ and $\overline{r}_{i}$. These
escape rates are sums over all the rates of transitions leaving a state, and the
different irreversible transitions must therefore
contribute additively to the escape rates. The result is that several
irreversible rates contribute additively to $\Sigma$, and the dynamic
contribution has the form
$\sum_{\alpha} R_{\alpha^+ \alpha^-} \left[ \tau_{\alpha^+} (\gamma) - \tau_{\alpha^-}
(\gamma)\right] $,
where $\alpha$ runs over different irreversible transitions, which connect
state $\alpha^-$
to state $\alpha^+$, and have the rate $R_{\alpha^+ \alpha^-} $.

The last generalization we consider is to systems with time dependent
rates. For such systems the conditional 
probability factors expressing the probability to stay in a state
 have the form $\exp \left[- \int_{t_{i} }^{t_{i+1}} dt r_{\alpha} (t) \right]$, 
in contrast to factors of $\exp \left[ - r_{\alpha} (t_{i+1}- t_{i}) \right]$ appearing in autonomous systems. The resulting
contribution to $\Sigma$ has dynamical terms
of the form $\int dt R_{\alpha^+ \alpha^-}  (t) \chi_{\alpha^-} (\gamma(t))$ replacing the terms
$R_{\alpha^+ \alpha^-}  \tau_{\alpha^-}$.

Based on these considerations the integral fluctuation theorem (\ref{eq:intft}) holds for time
dependent jump processes with several irreversible transitions, and, as long as the
probability distribution is non-vanishing, $\Sigma$ takes the form
\begin{equation}
\Sigma (\gamma) = \Delta S_{rev} (\gamma)+\sum_{\alpha} \int dt R_{\alpha^+ \alpha^-} (t) 
\left[ \chi_{\alpha^+} (\gamma(t)) - \chi_{\alpha^-}
(\gamma(t))\right].
\end{equation}

\section{Convergence of the exponential average}
\label{sec:conv}

Exponential averages, such as the one in Eqs. (\ref{eq:intft}) and
(\ref{eq:finalIFT}), often exhibit poor convergence. The underlying reason
is the difference between typical and dominant realizations. Typical
realizations are the ones which are likely during the
process of interest, and correspond to $\Sigma$ values in the vicinity of the
maximum of $P(\Sigma)$. In contrast,
the dominant realizations are those for which $e^{ -\Sigma} P(\Sigma)$ is
maximal. Jarzynski has discussed 
the convergence of exponential averages of this type using a gas in an expanding
piston as an example~\cite{Jarzynski2006}.
He used the detailed version of the fluctuation theorem to argue that the dominant realizations are actually the (time-reversed)
typical realizations of the
corresponding reversed process. In addition he has derived a simple estimate for
the number of realizations
needed for convergence of the exponential average. The purpose of this section
is to demonstrate that these
considerations also apply to systems with unidirectional transitions, and also
to numerically verify the validity of Eq. (\ref{eq:finalIFT}).

To do so we simulate the jump process of Sec.~\ref{sec:4states} using the
Gillespie algorithm. This algorithm efficiently generates stochastic trajectories with
the correct distribution by determining the time of the next transition, making
use of the fact that the waiting times between jumps are distributed exponentially~\cite{Gillespie1977}.
The transition rates were taken to be
$R_{12}=3$, $R_{21}=0.24$, $R_{23} =4$, $R_{32} =1$, $R_{34} =0.67$, $R_{43}
=2.1$, $R_{14} =1$, $R_{41} =0.78$ and $R_{42}=2.3$.
The jump process is assumed to be at steady state. For the parameters above we
find $p^{ss}(1) \simeq 0.5213$, $p^{ss}(2) \simeq 0.0515$,
$p^{ss}(3) \simeq 0.0498$ and $p^{ss}(4) \simeq 0.3772$.

\begin{figure}
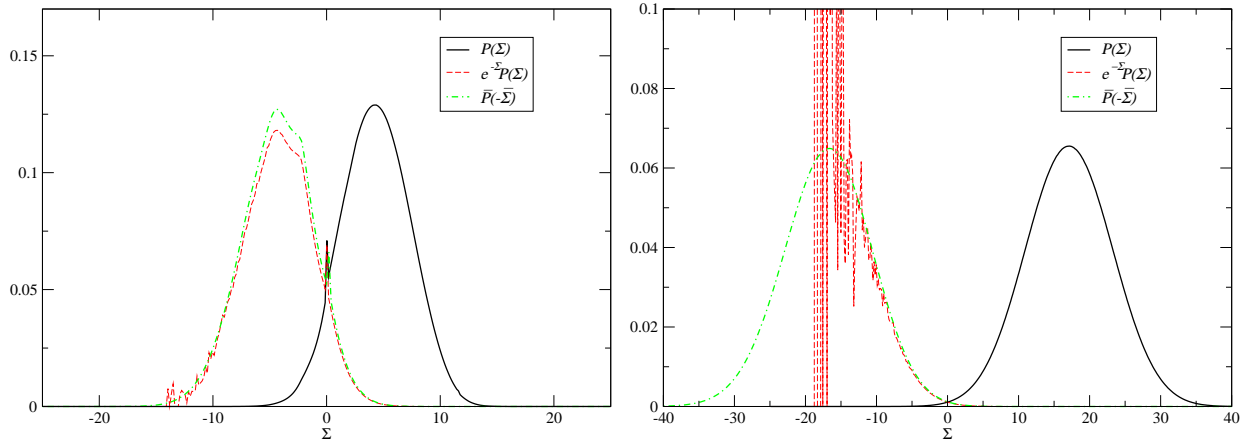

\centering
\mbox{\subfigure{\includegraphics[scale=0.35]{Fig2a.eps}}\hspace{.1cm}
\subfigure{\includegraphics[scale=0.35]{Fig2b.eps}}}
\caption{(Color online) The distribution of $\Sigma$ in the jump process
depicted in Fig.~\ref{fig:system} (black solid line). The dashed red line
depicts $ e^{ -\Sigma} P(\Sigma)$ calculated from $ P(\Sigma)$. The
dashed-dotted green line correspond to the distribution 
$\overline{P}(-\overline{\Sigma})$
obtained from simulations of the auxiliary dynamics. The results in the left
panel are for $t_f=5$, whereas the
right panel is calculated for $t_f=20$.}
 \label{fig-2}
\end{figure}

The numerically computed probability distribution of $\Sigma$ is depicted in the
left panel of Fig.~\ref{fig-2}
for $t_f=5$.
This distribution was generated from $10^9$ different realizations of the
process. This panel also depicts
$e^{ -\Sigma} P(\Sigma)$, which was calculated from $P(\Sigma)$. In addition it
shows the distribution
$\overline{P}(-\overline{\Sigma})$, which was calculated by numerical simulation
of the auxiliary dynamics with
the suitable initial condition ($p^{ss}$). The latter two curves are expected to
be identical. They are indeed very close to each other, and
the differences between them are possibly due to a combination of imperfect sampling of the tail
of $P(\Sigma)$ and of errors introduced
by binning the sparsely sampled region in the tail of $P(\Sigma)$. The spike at
$\Sigma=0$ is due to a discrete contribution to the probability density
from the 
trajectories which start at states $1$ or $3$ and never
make a jump during
the whole process. There are also discrete contributions
at $R_{42} t_f$ and $-R_{42} t_f$ from trajectories spending the whole time at states $4$ and $2$ respectfully.
The weights of these contributions is smaller compared to the one at $\Sigma=0$  because of the specific transition rates used
in the numerics. In contrast to these discrete features, trajectories which make jumps
lead to a continuous distribution due to the continuous nature of jump times.
Therefore the relative weight of discrete contributions to
$P(\Sigma)$ is reduced when $t_f$ is increased.
We note in passing that while the process we are interested in is stationary,
the corresponding
reference auxiliary is not stationary since $p^{ss} \neq \overline{p}^{ss}$.

The same curves are presented in the right panel of Fig.~\ref{fig-2} for
$t_f=20$. A comparison 
of the two panels shows that when $t_f$ is increased the dominant realizations
are pushed 
further into the tails of $P(\Sigma)$. This is easily seen from the amount of
overlap between
$P(\Sigma)$ and $\overline{P}(-\overline{\Sigma})$. Larger values of $t_f$ will
show even less overlap.

This reduced amount of overlap has a direct impact of the probability to
reliably obtain dominant realizations
with the correct weights, and hence on the convergence of exponential average
Eq. (\ref{eq:finalIFT}). For $t_f=5$ the dominant region
is well sampled by the simulation. In contrast, when $t_f=20$ it is clear that
the dominant region is only partially sampled, and the sampling is rather noisy.
In some spots 
the simulation of the original process gives weights which are too low or two
high. The leftmost part of the dominant region was never sampled. As a result
we expect that a simulation based calculation of $\left< e^{-\Sigma} \right>$
will give reasonable results for $t_f=5$
and somewhat poor results for $t_f=20$. Sampling of the latter can be improved
by adding more realizations.

\begin{figure}
\centering
\includegraphics[scale=0.6]{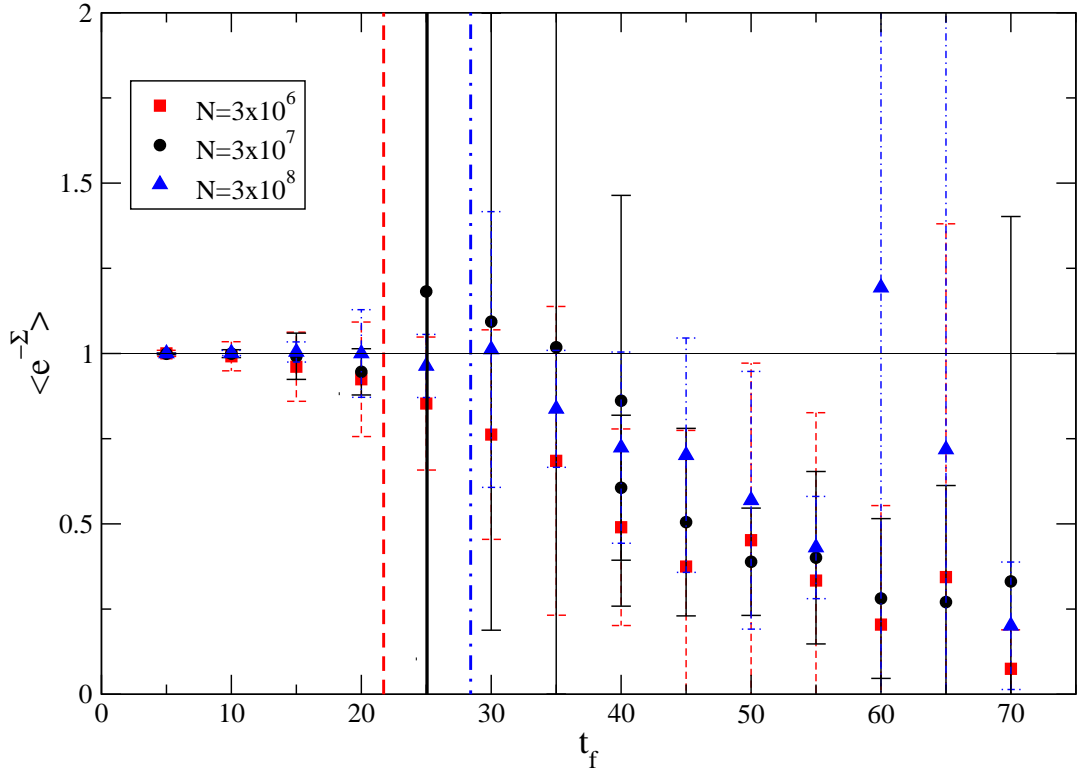}
\caption{(Color online) Numerical estimate of the exponential average
(\ref{eq:finalIFT}), obtained from
$N=3 \times 10^6$ (squares), $3 \times 10^7$ (circles) and $3 \times 10^8$
(triangles) realizations as a function of $t_f$. 
The error bars are estimates of the standard deviation obtained by dividing the
ensemble into 30 sub-ensembles and summing each one separately. The thick vertical lines
depict the approximate criterion for convergence, derived from 
Eqs. (\ref{eq:convest}) and (\ref{eq:esttime}), for  
$N=3 \times 10^6$ (red, dashed), $3 \times 10^7$ (solid, black) and $3 \times 10^8$
(dashed-dotted, blue).}
 \label{fig-3}
\end{figure}

To check the validity and convergence of the exponential average
(\ref{eq:finalIFT})
we have used ensembles of $3 \times 10^6$, $3 \times 10^7$ and $3
\times 10^8$
realizations of the jump process of Sec.~\ref{sec:4states}. Here $R_{42}=0.3$,
while the rest of the transition rates are identical to those used 
to generate Fig.~\ref{fig-2}. The results are presented in Fig.~\ref{fig-3}.
The ensemble was divided into $30$ sub-ensembles which were summed separately
and used to generated an effective standard deviation
measuring the fluctuations between different sub-ensembles.  It is clear that
good convergence is obtained for small $t_f$ where the average is close to $1$
and the standard deviation is small. When $t_f$
is increased the fluctuations between sub-ensembles become noticeable. When $t_f$ is increased even further,
and the dominant region is pushed further into the tail of $P(\Sigma)$, the dominant region is typically under-sampled and 
the numerical simulation returns an average value which is substantially smaller
than $1$. Occasionally, this region is over sampled and then sometimes the
simulation 
returns a value which can be larger than $1$. This is the hallmark of poorly
converged exponential averages of this type.

An estimate for the number of realizations needed for convergence was derived by
Jarzynski \cite{Jarzynski2006}. For the jump process studied here
it is given by
\begin{equation}
\label{eq:convest}
N^* \approx e^{\overline{\Sigma}_{typ} },
\end{equation}
where $\overline{\Sigma}_{typ}$ is the typical value of $\Sigma$ in the
auxiliary dynamics. This is an approximate criterion, and we
will further simplify it by estimating  $\overline{\Sigma}_{typ}$ as if the
auxiliary dynamics is at steady state. As was mentioned earlier this
is not true since the auxiliary dynamics exhibits transient relaxation towards its
steady state. We nevertheless make this approximation and obtain
\begin{equation}
\label{eq:esttime}
\overline{\Sigma}_{typ} \approx \left[ \overline{q}^{ss}_{rev}+R_{42} \left(
\overline{p}^{ss} (2) - \overline{p}^{ss} (4) \right) \right] t_f \simeq 0.6866
t_f.
\end{equation}
By substituting Eq.(\ref{eq:esttime}) in Eq. (\ref{eq:convest}) one can
calculate the value $t_f^*=\ln N /0.6866$ so that the exponential average converges 
for a given number of realizations as long as $t_f < t_f^*$. 
The three vertical lines (dashed, solid, and dash-dot) in Fig.~\ref{fig-3} represent 
$t_f^*$ for
$N=3 \times 10^6$, $3 \times 10^7$ and $3 \times 10^8$, respectively.
Since the estimate is approximate, and we further ignored the possible
contribution of transients, we only expect this estimate to work qualitatively.
Indeed all lines are 
roughly located in the transition region between times where the exponential
average converges and the region where it does not. Overall the numerical
results presented 
in Fig.~\ref{fig-3} qualitatively agree with our existing understanding of the
difficulties in numerical estimation of exponential averages. 

\section{Summary}
\label{sec:disc}

The stochastic coarse-grained dynamics of small systems in contact with external
thermal reservoirs exhibit microreversibility which
ultimately stems from the time reversal symmetry of the underlying deterministic
evolution. Nevertheless, there are situations
in which it is useful to consider models which violate microreversibility. 
We have studied the fluctuations of jump processes in a system with 
one or more unidirectional rates.
Such systems
violate microreversibility and can be viewed as motivated by physical processes
such as
spontaneous relaxation in quantum systems.
The usual formulation of fluctuation 
theorems can not be used for irreversible systems since it involves contributions 
to the entropy production of the
form $\ln \frac{R_{ji}}{R_{ij}}$ which are not always defined.

For such systems one can compare the dynamics to that of an auxiliary system in
which the unidirectional transitions are flipped.
However, this auxiliary dynamics has no simple physical interpretation, and
therefore it is of interest to identify measures of fluctuations which can be expresses
only in terms of the physical system.
We have shown that it is possible to derive such an integral
fluctuation,
Eq. (\ref{eq:finalIFT}), which has 
a simple and appealing physical interpretation. The realization dependent
 quantity, $\Sigma$,  which appears in the exponent of Eq.
(\ref{eq:finalIFT})
is a sum of a well defined entropy production 
due to bidirectional transitions and a dynamical term which
depends on the 
residence times in the states 
connected by the unidirectional transition. It can therefore be viewed as
including both thermodynamic and 
dynamical contributions. 

The validity of Eq. (\ref{eq:finalIFT}) was checked numerically using
simulations of the jump process. It is well known that exponential averages 
show poor convergence when dominant realizations are insufficiently sampled.
This was discussed in detail for systems
exhibiting microreversibility \cite{Jarzynski2006} and our numerical results
suggest that the same considerations can be applied also
for systems with unidirectional transitions. When the numerical results converge
they indeed support the validity of the integral 
fluctuation theorem.

\section*{Acknowledgments}
We thank Christopher Jarzynski and Massimiliano Esposito for illuminating
discussions.
SR is grateful for support from the Israel Science Foundation (grant 924/11) and
the US-Israel Binational Science Foundation
(grant 2010363). This work was partially supported by the COST Action MP1209. 
UH acknowledges support from the Indian Institute of Science, Bangalore, India.


\begin{thebibliography}{99}

\bibitem{Evans1993}
D. J. Evans, E. G. D. Cohen, and G. P. Morriss, {\em Phys. Rev. Lett.}, {\bf
71},  3616 (1993).

\bibitem{Evans1994}
D. J. Evans and D. J. Searles, {\em Phys. Rev. E}, {\bf 50},  1645 (1994).

\bibitem{Gallavotti1995}
G. Gallavotti and E. G. D. Cohen, {\em Phys. Rev. Lett.}, {\bf 74},  2694
(1995).

\bibitem{Kurchan1998}
J. Kurchan, {\em J. Phys. A: Math. Gen}, {\bf 31},  3719 (1998).

\bibitem{Lebowitz1999}
J. L. Lebowitz and H. Spohn, {\em J. Stat. Phys.}, {\bf 95},  333 (1999).

\bibitem{Rahav2013}
S. Rahav and C. Jarzynski, {\em New J. Phys.}, {\bf 15}, 125029 (2013).

\bibitem{Harbola2014}
U. Harbola, C. Van den Broeck, and K.Lindenberg, Phys. Rev. E {\bf 89}, 012141 (2014).

\bibitem{Seifert2005}
U. Seifert, {\em Phys. Rev. Lett.}, {\bf 95},  040602 (2005).

\bibitem{Jarzynski1997}
C. Jarzynski, {\em Phys. Rev. Lett.}, {\bf 78},  2690 (1997).

\bibitem{Crooks1998}
G. Crooks, {\em J. Stat. Phys}, {\bf 90},  1481 (1998).

\bibitem{Evans2002}
D. J. Evans and D. J. Searles, {\em Adv. Phys.}, {\bf 51},  1529 (2002).

\bibitem{Maes2003}
C. Maes and K. Net\v{o}cn\'{y}, {\em J. Stat. Phys.}, {\bf 110},  269 (2003).

\bibitem{Bustamante2005}
C. Bustamante, J. Liphardt, and F. Ritort, {\em Phys. Today.}, {\bf 58},  43
(2005).

\bibitem{Kurchan2007}
J. Kurchan, {\em J. Stat. Mech.}, {\bf 2007}  P07005.

\bibitem{Harris2007}
R. Harris and G. M. Sch\"{u}tz, {\em J. Stat. Mech.}, {\bf 2007}  P07020.

\bibitem{Imparato2007}
A. Imparato and L. Peliti, {\em C. R. Physique}, {\bf 8},  556 (2007).

\bibitem{Ritort2008}
F. Ritort, {\em Adv. Chem. Phys.}, {\bf 137},  31 (2008).

\bibitem{Esposito2009}
M. Esposito, U. Harbola, and S. Mukamel, {\em Rev. Mod. Phys.}, {\bf 81},  1665
(2009).

\bibitem{Broeck2010}
C. Van den Broeck, {\em J. Stat. Mech.}, {\bf 2010}, P10009.

\bibitem{Ciliberto2010}
S. Ciliberto, {\em J. Stat. Mech.}, {\bf 2010},  P12003.

\bibitem{Jarzynski2011}
C. Jarzynski, {\em Annu. Rev. Condens. Matter Phys.}, {\bf 2},  329 (2011).

\bibitem{Campisi2011}
M. Campisi, P. H\"{a}nggi, and P. Talkner, {\em Rev. Mod. Phys.}, {\bf 83}, 771
(2011).

\bibitem{Seifert2012}
U. Seifert, {\em Rep. Prog. Phys.}, {\bf 75},  126001 (2012).

\bibitem{Sekimoto1997}
K. Sekimoto, {\em J. Phys. Soc. Japan}, {\bf 66}, 1234 (1997).


\bibitem{Wang2014}
C. Wang and D. E. Feldman, {\em Int. J. of Mod. Phys. B}, {\bf 28}, 1430003
(2014).

\bibitem{Dorfman2012}
K. E. Dorfman, D. V. Voronine, S. Mukamel and M. O. Scully, {\em PNAS}, {\bf 110}, 2746
(2012).

\bibitem{Ting2009}
C. L. Ting, D. E. Makarov, and Z.-G. Wang, {\em J. Phys. Chem. B}, {\bf 113}, 4970
(2009).

\bibitem{Ohkubo2009}
J. Ohkubo, {\em J. Phys. Soc. Japan}, {\bf 78}, 123001
(2009).


\bibitem{Chong2010}
S-H. Chong, M. Otuski, and H. Hayakawa, {\em Phys. Rev. E}, {\bf 81}, 041130
(2010).

\bibitem{Avraham2011}
D. Ben-Avraham, S. Dorosz, and M. Pleimling, {\em Phys. Rev. E}, {\bf 84},
011115 (2011).

\bibitem{Zeraati2012}
S. Zeraati,  F. H. Jafarpour, and H. Hinrichsen, {\em J. Stat. Mech.}, {\bf
2012}  L12001.


\bibitem{Murashita2014}
Y. Murashita, K. Funo, and M. Ueda, preprint, arXiv:1401.4494 

\bibitem{Maes2008}
C. Maes, K. Net\v{o}cn\'{y}, and B. Wynants , {\em Markov Proc. Rel. Fields},
{\bf 14},  445 (2008).

\bibitem{Baiesi2009}
M. Baiesi, C. Maes, and B. Wynants, {\em Phys. Rev. Lett.}, {\bf 103},  010602
(2009).

\bibitem{Jarzynski2006}
C. Jarzynski, {\em Phys. Rev. E}, {\bf 73}, 046105 (2006).

\bibitem{Gillespie1977}
D. T. Gillespie, {\em J. Phys. Chem.}, {\bf 81}, 2340 (1977).

\end{thebibliography}
\end{document}